\documentclass[english]{article}
\usepackage{graphicx}
\def\no{\noindent}
\def\bc{\begin{center}}
\def\ec{\end{center}}

\def\beq{\begin{equation}}
\def\eeq{\end{equation}}

\begin{document}

\title{Long-range correlations in disordered graphene
}

\author{K. Ziegler\\
Institut f\"ur Physik, Universit\"at Augsburg\\
D-86135 Augsburg, Germany}
\date{\today}

\maketitle

Abstract:

The appearence of long-range correlations near the Dirac point of a
Dirac-like spinor model with random vector potential is studied. These
correlations originate from a spontaneously broken symmetry and their
corresponding Goldstone modes. Using a strong-disorder expansion,
correlation functions and matrix elements are analyzed and compared with
results from a weak-disorder expansion. The local 
density of states correlation and the overlap between states above and
below the Dirac point are characterized by a long-range behavior.
The correlation range decreases with the distance
from the Dirac point. Transport is diffusive
and the diffusion coefficient is proportional
to the one-particle scattering time for any strength of disorder. 
A consequence of the special properties of
particle-hole scattering is a constant microwave conductivity for 
weak as well as for strong disorder, describing a 
deviation from conventional Drude-like transport. 
Some properties of the model can be linked to a 
kind of Kondo scale, which is generated by disorder. Finally, the properties
of the wave functions at the Dirac point are characterized by their 
participation ratios, indicating a critical state at the Dirac point. 


\section{Introduction}

The electronic properties of graphene are closely related to the existence
of a spinor wave function and the two Dirac nodes in the band structure 
\cite{novoselov05,zhang05,geim07}. 
This implies an unconventional behavior which is associated with the Klein paradox
\cite{katsnelson06a}. An important effect is the scattering between the 
hole and the particle sector of the Dirac cones, leading, for instance,
to the zitterbewegung 
\cite{cserti06,katsnelson06,katsnelson07a,rusin07,zuelicke07,schliemann08}. 
A constant contribution to the microwave
conductivity is another consequence of this effect, causing a 
deviation from conventional Drude-like transport 
\cite{ziegler07,stauber08,nair08}. The latter has
been studied only for weak disorder. It would be interesting to analyze these effects,
their modification or even their destruction in the presence of strong disorder.

The presence of disorder in graphene has been discussed in a number of recent papers,
considering different physical conditions. It seems that disorder should 
appear in the effective Dirac-like Hamiltonian in the form of a random vector potential 
due to instability of the translational order in 2D 
\cite{morozov06,geim07,castroneto07b,katsnelson07aa,guinea08}. 
Scattering by a random vector potential affects only the phase of the wave functions.
Its influence on transport is such that no localization has been observed, even for very strong
disorder \cite{ziegler08}. However,
it is believed that disorder has a substantial effect on the magnetoresistance, observed
as a suppression of the weak-localization peak \cite{morozov06,cho08}. Another
effect, which might be caused by disorder, are fluctuations of the charge distribution 
near the Dirac point. Recent experiments have revealed that in graphene long-range 
charge fluctuations exist \cite{martin07}. 

The Dirac point is very special because it is associated with zero-energy modes and
a vanishing density of states. Consequently, it might be very sensitive to disorder. It is known
that, depending on the type of disorder, the properties of Dirac fermions are strongly
affected by randomness \cite{ludwig94,altland02}. An example is the prediction of a dramatic change of the 
density of states due to disorder. On the other hand, some quantities, like the conductivity,
are remarkably robust \cite{fradkin86,ludwig94,ziegler97,ziegler06}.  
Of particular interest is the question how disorder affects the
interband scattering. Weak disorder has no effect on particle-hole scattering, only on the inter-particle
scattering (cf. \cite{ziegler07}). All this indicates
that the Dirac point is an interesting object to study 
in more detail, especially when disorder is strong.

To understand the effect of a random vector potential in a Dirac-like Hamiltonian, 
it has to be noticed that the Dirac point is characterized by a chiral symmetry. 
We discuss this symmetry and its spontaneous breaking by the random vector potential. 
It is known that the spontaneously broken symmetry implies a robust diffusion mode
for weak disorder \cite{ziegler97} and for strong disorder \cite{ziegler08}. 
In this paper we will demonstrate that the local density of states correlations
have a long-ranged behavior, although the average density of states is finite.
Some properties can be linked to a kind of Kondo scale, generated by disorder 
\cite{auslender07,dora08}. 
The analysis of the scattering between different parts of the Dirac cone leads
to correlations between different energy states, whose correlation are expressed
in terms of transition matrix elements. A special case is the corresponding 
matrix elements of the position operator with respect to $\pm E$ energy states, which
appear as important contributions to the microwave conductivity.
Finally, the second moment of the 
local density of states can be associated with the inverse participation ratio. 
This quantity characterizes the properties of the wave functions at the Dirac point.

In order to study the effect of symmetry breaking in the presence of disorder, we
apply a perturbation theory for strong disorder. Strong-disorder expansions have been
quite successful in different fields of physics. In the case of Dirac fermions
it has been observed that the expansion has a number of advantages in comparison
with a weak-disorder expansion \cite{ziegler08}. The main reason is that the leading
order of the expansion is quite simple. Nevertheless, the strong-disorder expansion
cannot be applied directly to the one-particle Green's function of the Dirac-like 
Hamiltonian because it
has, in contrast to a Schr\"odinger-like Hamiltonian, divergent contributions from 
self-crossing loops.  (The strong-disorder expansion can be formulated as an expansion
in terms of the hopping elements of the Hamiltonian.) These problems can be avoided by
applying the expansion to the saddle-point integration. Then the integration
is restricted to the most important contributions, namely the saddle-point manifold,
that takes into account the underlying symmetries of the model. In this case the 
strong-coupling expansion can be controlled and leads in
leading order to a model with diffusion propagator. It describes the Goldstone
modes which are generated by spontaneous symmetry breaking. The results of this
expansion can be compared with those of the more common weak-disorder expansion.
The latter gives in leading order a factorization of the average two-particle 
Green's function into a product of two average one-particle Green's functions,
which can be treated within the self-consistent Born approximation. Both approaches
are compared to study similarities and differences of both regimes.

The paper is organized as follows. In Sect. 2 we introduce the generalized Dirac model
and discuss spontaneous symmetry breaking. This includes the definition of some physical 
quantities, which are studied in the rest of the paper. The framework of functional
integration for our model is presented in Sect. 3, and the corresponding symmetries
are analyzed in Sec. 4. Then the averaging with respect to the random vector potential
and the related saddle-point integration are performed (Sect. 5). The results of the
calculations are discussed and compared with weak-disorder approximations
in Sect. 6.

\section{Model and spontaneous symmetry breaking}

In this paper we study the generalized Dirac Hamiltonian for a spinor-1/2 state
in a random vector potential
\beq
H=H_0+v\sigma_1 ,
\hskip0.5cm
H_0=h_1\sigma_1+h_2\sigma_2 \ .
\label{ham000}
\eeq
$\sigma_j$ ($j=0,1,2,3$) are Pauli matrices,
$h_j$ is an operator that acts in space, and $v$ is space-diagonal with
random variables $v_r$. The latter are independently Gaussian distributed with zero mean
and variance $g$.
Special cases for $H_0$ are 2D Dirac fermions, whose Fourier components are
\[
h_j=k_j
\]
and the tight-binding model on a honeycomb lattice with
\[
h_1=-t\sum_{j=1}^3\cos(\vec{a}_j\cdot\vec{k}) ,
\hskip0.5cm
h_2=-t\sum_{j=1}^3\sin(\vec{a}_j\cdot\vec{k})
\]
with the lattice vectors of the honeycomb lattice
\[
\vec{a}_{1}=(-\sqrt{3}/2,1/2),
\hskip0.5cm
\vec{a}_{2}=(0,-1),
\hskip0.5cm
\vec{a}_{3}=(\sqrt{3}/2,1/2). 
\]
$H_0$ can be diagonalized 
as $H_0=diag(e_k,-e_k)$ with $e_k=\sqrt{h^2_1+h^2_2}$. 
The Hamiltonian $H$ is invariant under the continuous
transformation $H\to e^{i\alpha\sigma_3}He^{i\alpha\sigma_3}$. This symmetry can be
broken spontaneously. To measure spontaneous symmetry breaking (SSB), we consider the
one-particle Green's function
\[
G(z)=(H+z)^{-1}
\]
and calculate the difference of this expression before and after the symmetry
transformation. For the special case with parameter value $\alpha=i\pi/2$ this reads
\[
G(z)-i\sigma_3G(z)i\sigma_3=(H+z)^{-1}-(H-z)^{-1} \ .
\]
Since $z$ breaks the symmetry, we send it to zero. If we choose $z=i\epsilon$, the
results for the diagonal elements of $G$ are proportional to the local density of states at the
Dirac point:
\[
\rho_r(0)=\delta_{rr}(H)
\propto\lim_{\epsilon\to0}[(H+i\epsilon)^{-1}_{rr}-(H-i\epsilon)^{-1}_{rr}] \ .
\]
Thus, a nonzero $\rho_r(0)$ indicates SSB. 
This is similar to spontaneous symmetry breaking in classical systems, e.g., in the
case of the magnetic phase transition of a classical Heisenberg model \cite{fradkin91}.
However, there is a remarkable difference: In the Heisenberg model there is SSB only
for systems with dimensionality $d$ higher than 2, whereas this restriction does not
hold in a disordered system. Even for $d=0$, where $H=v\sigma_1$, we have SSB, since the DOS reads
\[
\langle\rho(0)\rangle=\lim_{\epsilon\to0}\int\frac{\epsilon}{\epsilon^2+v^2}P(v)dv 
=\lim_{\epsilon\to0}\int\frac{1}{1+x^2}P(\epsilon x)dx >0
\] 
for a continuous distribution function $P(v)$, e.g. a Gaussian. Another interesting point is
that the second moment of the density of states diverges like $\epsilon^{-1}$:
\beq
\langle\rho(0)^2\rangle=\int\frac{\epsilon^2}{(\epsilon^2+v^2)^2}P(v)dv
=\epsilon^{-1}\int\frac{1}{(1+x^2)^2}P(\epsilon x)dx \ .
\label{2moment}
\eeq
This simply reflects the fact that the distribution of the density of states has a
Lorenzian form.

\subsection{Physical quantities}

A characteristic property of the Hamiltonian $H$ in Eq. (\ref{ham000}) is that its
spectrum is symmetric with respect to the energy $E=0$ (the Dirac point), consisting
of states $\Psi_{\pm E}$ in the upper and the lower band (particle and hole parts of the Dirac cones). 
These states are connected by the transformation $\Psi_{-E}=\sigma_3\Psi_{E}$ because the eigenvalue
equation
\[
H\Psi_E=E\Psi_E
\]
gives
\[
\sigma_3H\sigma_3\sigma_3\Psi_E=E\sigma_3\Psi_E
\]
which becomes with $\sigma_3H\sigma_3=-H$
\[
H\sigma_3\Psi_E=-E\sigma_3\Psi_E \ .
\]
Scattering between $\Psi_{E}$ and $\Psi_{-E}$ leads to interesting physical phenomena, 
like the zitterbewegung \cite{cserti06,katsnelson06} 
or a constant contribution to the microwave conductivity 
\cite{ziegler07,mikhailov07,nair08}. The latter can be seen by considering a typical contribution
to the frequency-dependent conductivity at frequency $\omega$ (cf. Appendix A) as
\[
\sigma^{II}(\omega)
=-\frac{e^2}{4h}\omega\int_{-\omega/2}^{\omega/2}
\sum_{r}r_k^2 Tr_2[G_{r0}(\omega/2-E-i\epsilon)G_{0r}(-\omega/2-E+i\epsilon)
\]
\beq
+G_{r0}(-\omega/2+E-i\epsilon)G_{0r}(\omega/2+E+i\epsilon)]dE \ ,
\label{cond0a}
\eeq
where $Tr_2$ is the trace with respect to Pauli matrices.
This can also be expressed in terms of the matrix elements of $r_k^2$ between the states 
$\Phi_{\pm\omega/2-E}$ (cf. Appendix B) as
\beq
\sigma^{II}(\omega)
=-\frac{e^2}{4h}\omega\int_{-\omega/2}^{\omega/2}
(\langle  \Phi_{-\omega/2+E}|r_k^2|\Phi_{\omega/2+E}\rangle 
+\langle  \Phi_{\omega/2-E}|r_k^2|\Phi_{-\omega/2-E}\rangle )dE \ .
\label{cond00b}
\eeq
For a system at the Dirac point $E=0$, we expect that the main contribution to the conductivity
comes from the scattering between $\Phi_{-\omega/2}$ and $\Phi_{\omega/2}$, where the scattering does
not change momentum (cf. Fig. 1). Therefore, we consider in the following only
\beq
\sigma_0(\omega)
=-\frac{e^2}{4h}\omega^2\langle  \Phi_{-\omega/2}|r_k^2|\Phi_{\omega/2}\rangle 
+\langle  \Phi_{\omega/2}|r_k^2|\Phi_{-\omega/2}\rangle ) 
=-\frac{e^2}{2h}\omega^2 \langle  \Phi_{-\omega/2}|r_k^2|\Phi_{\omega/2}\rangle \ ,
\label{cond0b}
\eeq
where the last equation is a consequence of the symmetry of the matrix element.
Then a constant conductivity, found theoretically \cite{ziegler07,stauber08} and 
experimentally \cite{nair08},
requires a matrix element that diverges for $\omega\sim0$ as
\[
\langle  \Phi_{-\omega/2}|r_k^2|\Phi_{\omega/2}\rangle\sim\omega^{-2} \ .
\]

There are two other correlation functions that are of interest for the characterization
of the disordered system of quasiparticles in graphene. One is
\beq
C_{rr'}=\langle Tr_2[\delta_{rr'}(H-E)\delta_{r'r}(H-E)]\rangle
\label{corr00b}
\eeq
which is related to the DC conductivity $\sigma_{\mu\mu}$ at $T=\omega=0$ through the 
expression \cite{ziegler06}
\beq
\sigma_{\mu\mu}=2\pi\frac{e^2}{\hbar}\lim_{\epsilon\to0}
\epsilon^2\sum_{r'}(r_\mu-r_\mu')^2C_{rr'} \ .
\label{cond000}
\eeq
The other correlation function is
\beq
D_{rr'}=\langle Tr_2[\delta_{rr}(H-E)]Tr_2[\delta_{r'r'}(H-E)]\rangle
\label{corr00c}
\eeq
which is the correlation of the local density of states at sites $r$ and $r'$. 
For $r'=r$, $D_{rr}$ is related to the participation ratio and the inverse 
participation ratio (cf. Appendix C). These are quantities that measure the 
statistical properties of the wavefunctions and their localization behavior \cite{wegner80}. 
The participation ratio
\[
p^{(2)}=\frac{\langle \rho_r\rangle^2}{\langle \rho_r^2\rangle}
\]
vanishes for localized states because the local density of states $\rho_r$ has a broad
distribution with divergent second moments in this case, similar to 
the single-site density of states of Eq. (\ref{2moment}). Using $D_{rr}$ the participation ratio reads
\beq
p^{(2)}=\lim_{\epsilon\to0}\frac{\langle \rho_r\rangle^2}{D_{rr}} \ .
\label{pr1}
\eeq
Moreover, the inverse participation ratio is also related to the second moment of the local 
density of states (cf. Appendix C):
\beq
P^{(2)}=\lim_{\epsilon\to0}\epsilon D_{rr} \ .
\label{ipr1}
\eeq
It vanishes as one approaches the regime of extended states, coming from the
localized regime.

The density operator $\delta(H)$ can be expressed by the one-particle Green's function
\[
G^\pm =(H\pm i\epsilon)^{-1} 
\]
as
\[
\delta_{rr'}(H)=\frac{1}{2i\pi}(G^-_{rr'}-G^+_{rr'}) \ .
\]
With the relation $\sigma_3G^\pm\sigma_3=-G^\mp$ the delta function also reads
\[
\delta_{rr'}(H)=\frac{i}{2\pi}[G^+_{rr'}+\sigma_3G^+_{rr'}\sigma_3] 
=\frac{i}{\pi}\pmatrix{
G^+_{rr',11} & 0 \cr
0 & G^+_{rr',22} \cr
} \ .
\]
This relation implies for the correlation functions at the Dirac point $E=0$
\beq
C_{rr'}=-\frac{1}{\pi^2}\sum_{j=1,2}\langle G^+_{rr',jj}G^+_{r'r,jj}\rangle \ ,
\label{corr0a}
\eeq
\beq
D_{rr'}=-\frac{1}{\pi^2}\langle Tr_2[G^+_{rr}]Tr_2[G^+_{r'r'}]\rangle \ .
\label{corr0b}
\eeq
Moreover, away from the Dirac point at energies $E=\pm\omega/2$ it gives
\beq
A_{rr'}(\omega)=-\langle Tr_2\left[\sigma_3G_{rr'}(\omega/2+i\epsilon)
\sigma_3G_{r'r}(\omega/2+i\epsilon)\right]\rangle  \ ,
\label{corr0c}
\eeq
which can be used to write for the conductivity of Eq. (\ref{cond0b})
\beq
\sigma_0(\omega)=-\frac{e^2}{2h}\omega^2\sum_{r}r_k^2 A_{r0}(\omega) 
=-\frac{e^2}{2h}\omega^2\langle\Phi_{\omega/2} |r_k^2|\Phi_{-\omega/2}\rangle \; .
\label{conda1}
\eeq

\section{Functional Integral}

All three correlation functions in Eqs. (\ref{corr0a}), (\ref{corr0b}) and (\ref{corr0c})
are written as products of two Green's functions.
We can express these products, before averaging, as a Gaussian functional integral with
two independent Gaussian fields, a boson (complex) field $\chi_{rk}$ and a fermion 
(Grassmann) field $\Psi_{rk}$ ($k=1,2$) and their conjugate counterparts ${\bar\chi}_{rk}$
and ${\bar\Psi}_{rk}$:
\beq
G_{rr',jj'}(i\epsilon)G_{r'r,k'k}(i\epsilon)
=\int\chi_{r'j'}{\bar \chi}_{rj} \Psi_{rk} {\bar \Psi}_{r'k'}\exp(-S_0(E))
{\cal D}\Psi {\cal D}\chi \ ,
\label{finta}
\eeq
where $S_0(E)$ is a quadratic form of the four-component field 
$\phi_r=(\chi_{r1},\chi_{r1},\Psi_{r2},\Psi_{r2})$
\begin{equation}
S_0(E)=
-i
\sum_{r,r'}
\phi_r\cdot({\bf H}+i\epsilon+E)_{r,r'}{\bar\phi}_{r'} \ \ (\epsilon>0)\ .
\label{ssa0}
\end{equation}
The use of the mixed field $\phi_r$ has the advantage that an extra normalization
factor for the integral is avoided. 
The extended Hamiltonian ${\bf H}=diag(H,H)$ of $S_0$ acts 
in the boson and in the fermion sector separately. 
Using Eq. (\ref{finta}), the correlation functions now read
\[
C_{rr'}=-\frac{1}{\pi^2}\sum_j\int\chi_{r'j}{\bar \chi}_{rj} \Psi_{rj} {\bar \Psi}_{r'j}
\langle\exp(-S_0(0))\rangle
{\cal D}\Psi {\cal D}\chi \ ,
\]
\[
A_{rr'}(\omega)=-\sum_{j,j'}(-1)^{j+j'}\int\chi_{r'j'}{\bar \chi}_{rj} \Psi_{rj} {\bar \Psi}_{r'j'}
\langle\exp(-S_0(\omega/2)\rangle
{\cal D}\Psi {\cal D}\chi
\]
and
\[
D_{rr'}=-\frac{1}{\pi^2}\sum_{j,j'}\int\chi_{rj}{\bar \chi}_{rj} \Psi_{r'j'} {\bar \Psi}_{r'j'}
\langle\exp(-S_0(0))\rangle
{\cal D}\Psi {\cal D}\chi \ .
\]
In the case of $C_{rr'}$ and $A_{rr'}(\omega)$ we arrange the products of the fields such that pairs
at the same site are neighboring factors:
\[
C_{rr'}=\frac{1}{\pi^2}\sum_j\int\chi_{r'j}{\bar \Psi}_{r'j}\Psi_{rj}{\bar \chi}_{rj}
\langle\exp(-S_0(0))\rangle{\cal D}\Psi {\cal D}\chi \ ,
\]
\[
A_{rr'}(\omega)=\sum_{j,j'}(-1)^{j+j'}\int\chi_{r'j'}{\bar \Psi}_{r'j'}\Psi_{rj}{\bar \chi}_{rj}
\langle\exp(-S_0(\omega/2))\rangle{\cal D}\Psi {\cal D}\chi \ .
\]

\section{Supersymmetry}

The Hamiltonian ${\bf H}$ in $S_0$ is invariant under the transformation 
\beq
{\bf U}=U_0(q,p){\bf U}_S U_0(q',p')=\pmatrix{
e^{i(q+q')\sigma_3}(1+\psi{\bar\psi}/2) & e^{i(q+p')\sigma_3}\psi\sigma_3 \cr
e^{i(q'+p)\sigma_3}{\bar\psi}\sigma_3 &  e^{i(p+p')\sigma_3}(1-\psi{\bar\psi}/2) \cr
}
\label{u0}
\eeq
with
\[
{\bf U}_S=\exp\pmatrix{
0 & \psi\sigma_3\cr
{\bar\psi}\sigma_3 & 0 \cr
} , \ \ 
U_0(q,p)=\pmatrix{
e^{iq\sigma_3} & 0 \cr
0 & e^{ip\sigma_3} \cr
} \ .
\]
For $\gamma_j=diag(\sigma_j,\sigma_j)$ there is the symmetry transformation 
\[
{\bf U}\gamma_j{\bf U}'={\bf U}{\bf U}^{-1}\gamma_j =\gamma_j
\]
with
\beq
{\bf U}'=U_0(q',p'){\bf U}_S U_0(q,p)=\pmatrix{
e^{i(q+q')\sigma_3}(1+\psi{\bar\psi}/2) & e^{i(q'+p)\sigma_3}\psi\sigma_3 \cr
e^{i(q+p')\sigma_3}{\bar\psi}\sigma_3 &  e^{i(p+p')\sigma_3}(1-\psi{\bar\psi}/2) \cr
} \ .
\label{u1}
\eeq
This implies the invariance
\beq
{\bf H}\to {\bf U}{\bf H}{\bf U}'={\bf H} \ .
\label{symm0}
\eeq
To simplify the notation one can introduce the phases $\varphi_{ij}$ with
\[
\varphi_{11}=q+q' ,
\hskip0.4cm
\varphi_{12}=q+p' ,
\hskip0.4cm
\varphi_{21}=q'+p ,
\hskip0.4cm
\varphi_{22}=p+p'
\]
and write
\beq
{\bf U}{\bf U}'=\pmatrix{
e^{2i\varphi_{11}\sigma_3}(1+\psi{\bar\psi}) +e^{2i\varphi_{12}\sigma_3}
\psi{\bar\psi} 
& [e^{i(\varphi_{12}+\varphi_{22})\sigma_3} + e^{i(\varphi_{11}+\varphi_{21})\sigma_3}]
\psi\sigma_3 \cr
[e^{i(\varphi_{11}+\varphi_{21})\sigma_3} + e^{i(\varphi_{12}+\varphi_{22})\sigma_3}]
{\bar\psi}\sigma_3
& e^{2i\varphi_{22}\sigma_3}(1-\psi{\bar\psi}) -e^{2i\varphi_{21}\sigma_3}
\psi{\bar\psi} \cr
} \ .
\label{fullsymm}
\eeq

\section{Averaged correlation functions}

Averaging Eq. (\ref{finta}) over the Gaussian distribution of $v_r$ means
replacing $\exp(-S_0)$ by $\langle \exp(-S_0)\rangle$ on the right-hand side of 
the equation. The latter can be written again as an exponential function
$\langle \exp(-S_0)\rangle=\exp(-S_1)$, where the new function $S_1$ contains also
quartic terms of the field $\phi$:
\beq
S_1=-i\sum_{r,r'}\phi_r\cdot({\bf H}_0+i\epsilon+E)_{r,r'}{\bar\phi}_{r'}
+g\sum_r (\phi_r\cdot\gamma_1{\bar\phi}_r)^2 \ .
\label{effaction}
\eeq
Then it is convenient to transform the integration variables (Hubbard-Stratonovich transformation) as 
\beq
\pmatrix{\chi_r{\bar\chi}_r&\chi_r{\bar\Psi}_r\cr
\Psi_r{\bar\chi}_r&\Psi_r{\bar\Psi}_r\cr
}\rightarrow
{\bf Q}_r=\pmatrix{
Q_r & \Theta_r\cr
{\bar\Theta}_r & -iP_r\cr
} \ ,
\eeq
where $Q_{r}$, $P_{r}$ are symmetric $2\times2$ matrices 
and $\Theta_{r}$, ${\bar\Theta}_{r}$
are $2\times2$ matrices whose elements are independent Grassmann variables. 
Now the correlation functions can be rewritten as correlation functions in the new field ${\bf Q}_r$: 
\beq
C_{rr'}=\frac{1}{g^2\pi^2}\sum_j\int (\sigma_1\Theta)_{r',jj}(\sigma_1{\bar\Theta})_{r,jj}
\exp(-S_2){\cal D}[{\bf Q}]
\label{corr2a}
\eeq
\beq
A_{rr'}(\omega)=\frac{1}{g^2}
\int Tr_2(\sigma_3\sigma_1\Theta_{r'})Tr_2(\sigma_3\sigma_1{\bar\Theta}_{r})\exp(-S_2){\cal D}[{\bf Q}]
\label{corr2b}
\eeq
and
\beq
D_{rr'}=-\frac{1}{g^2\pi^2}\int Tr_2(-i\sigma_1P_{r'})Tr_2(\sigma_1Q_r)
\exp(-S_2){\cal D}[{\bf Q}]
\label{corr2c}
\eeq
with
\beq
S_2=\sum_{r,r'}\frac{1}{g}{\rm Trg}({\bf Q}_{r}^2) 
+\ln [{\rm detg}[{\bf H}_0+i\epsilon+E -2\gamma_1{\bf Q}]] \ .
\label{action2}
\eeq
${\rm Trg}$ is the graded trace, ${\rm Tr}_2$ the trace with respect to Pauli matrices,
and ${\rm detg}$ the graded determinant \cite{ziegler97}.

\subsection{Saddle-point manifold}

The integration in Eqs. (\ref{corr2a}), (\ref{corr2b}), and (\ref{corr2c}) can be
performed in saddle-point approximation. The saddle point (SP) is obtained as the
solution of $\delta S_2=0$. Assuming a solution of the form
\[
{\bf Q}_0=-i\frac{\eta}{2}\gamma_1 \ ,
\]
we obtain the parameter $\eta$ from the SP equation
\beq
\eta=igTr_2G_{rr}(E+i\epsilon+i\eta) \ .
\label{SPE00}
\eeq
For $\epsilon=E=0$ the SP equation is invariant under the global symmetry 
transformation $\gamma_1{\bf Q}_0\to{\bf U}\gamma_1{\bf Q}_0{\bf U}'$ of Eq. (\ref{symm0}).
This transformation leads to the SP manifold
\beq
{\bf Q}_r'=-i\frac{\eta}{2}\gamma_1{\bf U}_r{\bf U}_r' 
=-i\frac{\eta}{2}{{\bf U}'}^{-1}_r\gamma_1{\bf U}'_r \ ,
\label{spm}
\eeq
where ${\bf U}_r$ and ${\bf U}_r'$ are obtained from Eqs. (\ref{u0}) and (\ref{u1}) 
by replacing the transformation parameters $\psi$, ...  by space-dependent
variables  $\psi_r$, ... The form of ${\bf Q}_r'$, which is dictated by the symmetry of Sect. 4,
implies for the action on the SP manifold that (i) the quadratic term
in $S_2$ vanishes and (ii) the remaining term becomes
\beq
S'=\ln {\rm detg}({\bf H}_0+i\epsilon+E+i\eta{\bf U}{\bf U}')
=\ln {\rm detg}({\bf U}^{-1}({\bf H}_0+i\epsilon+E){{\bf U}'}^{-1}+i\eta) \ .
\label{spaction}
\eeq
This action contains the symmetry breaking field $i\epsilon+E$.
Expansion in powers of $\eta^{-1}$ yields, after renaming ${\bf U}^{-1}\to {\bf U}$,
\[
S'=\frac{\epsilon-iE}{\eta}Trg({\bf U}{\bf U}')
+\frac{1}{\eta^2}Trg({\bf U}'{\bf U}{\bf H}_0{\bf U}'{\bf U}{\bf H}_0) + O(\eta^{-3}) \ .
\]
Rescaling $\varphi\to\eta^{-1}\varphi$ and $\psi\to\eta^{-1}\psi$ does not change the integration
measure and allows us to perform an expansion of ${\bf U}{\bf U}'$ in powers of $\eta^{-1}$
up to $o(\eta^{-3})$:
\beq
{\bf U}{\bf U}'=\pmatrix{
\sigma_0+2i\varphi_{11}\sigma_3/\eta -2\varphi_{11}^2\sigma_0/\eta^2+2\psi{\bar\psi}\sigma_0/\eta^2 
& 2\psi\sigma_3/\eta +2i(\varphi_{11}+\varphi_{22})\psi\sigma_0/\eta^2 \cr
2{\bar\psi}\sigma_3/\eta+2i(\varphi_{11}+\varphi_{22}){\bar\psi}\sigma_0/\eta^2
& \sigma_0+ 2i\varphi_{22}\sigma_3/\eta-2\varphi_{22}^2\sigma_0/\eta^2-2\psi{\bar\psi}\sigma_0/\eta^2 \cr
} \ . 
\label{approxfield}
\eeq
where we have used $\varphi_{12}+\varphi_{21}=q+p'+q'+p=\varphi_{11}+\varphi_{22}$.
This provides an expansion of the action, where the leading order is a quadratic form in terms of the fields
$\psi$ and $\varphi$:
\[
S'=\sum_{r,r'}K_{rr'}(\epsilon-iE)
(-\varphi_{11,r}\varphi_{11,r'}+\varphi_{22,r}\varphi_{22,r'}+2\psi_r{\bar\psi}_{r'})
\]
\beq
+\frac{2}{\eta^2}\sum_{r,r'}K_{rr'}(0)
(\varphi_{11,r}+\varphi_{22,r})(\varphi_{11,r'}+\varphi_{22,r'})\psi_r{\bar\psi}_{r'}
+O(\eta^{-7})
\label{faction}
\eeq
with
\beq
K_{rr'}(\epsilon-iE)=4\eta^{-4}\left[[\sum_{j=1,2}\sum_{{\bar r}}h_{j,r{\bar r}}h_{j,{\bar r}r}
+2(\epsilon-iE)\eta  
]\delta_{rr'}-\sum_{j=1,2}h_{j,rr'}h_{j,r'r}\right] 
\ .
\label{cmatrix}
\eeq
There is only one term that couples the Grassmann field $\psi$ with the field $\varphi$.
It turns out (cf. Appendix D) that this term drops out after the integration over $\varphi$.
Moreover, the Jacobian $J$ of the transformation should be $J=2^{-N}+o(\eta^{-2})$ ($N$ is the number of lattice sites)
in order to satisfy the condition
\beq
\int e^{-S'}J{\cal D}[\psi,\varphi]=1 \ .
\label{zunity}
\eeq
The symmetry-breaking term $\epsilon-iE$ appears as a prefactor of a diagonal term.
This action can now be used to calculate the correlation functions in Eqs. (\ref{corr2a}),
(\ref{corr2b}), and (\ref{corr2c}).
An expansion of the components of the matrix field yields
\[
\sigma_1Q =-\frac{i}{2}[
\eta\sigma_0+2i\varphi_{11}\sigma_3 +\frac{2}{\eta}(-\varphi_{11}^2+\psi{\bar\psi})\sigma_0] 
\]
\[
-i\sigma_1P =-\frac{i}{2}[
\eta\sigma_0+2i\varphi_{22}\sigma_3 -\frac{2}{\eta}(\varphi_{22}^2+\psi{\bar\psi})\sigma_0] 
\]
\[
\sigma_1\Theta=-i[
\sigma_3 +\frac{i}{\eta}(\varphi_{11}+\varphi_{22})\sigma_0)]\psi
\]
\[
\sigma_1{\bar\Theta}=-i[
\sigma_3 +\frac{i}{\eta}(\varphi_{11}+\varphi_{22})\sigma_0)]
{\bar\psi} \ .
\]
Integration over the fields $\psi$ and $\varphi$ with respect to the quadratic action $S'$ in
\beq
\langle ... \rangle =\int ... e^{-S'}{\cal D}[\psi,\varphi]
\label{spmint}
\eeq
leads to (up to $o(\eta^{-1})$)
\beq
\langle\sigma_1Q\rangle=i\sigma_0\eta/2, \ \ 
-i\langle\sigma_1P\rangle=i\sigma_0\eta/2 \ ,
\label{single}
\eeq
since the integration rules imply
\[
\langle-\varphi_{11}^2\rangle +\langle\psi{\bar\psi}\rangle
=\langle\varphi_{22}^2\rangle+\langle\psi{\bar\psi}\rangle=0 \ .
\]
It is also a consequence of Eq. (\ref{zunity}). 
Moreover, we need for the evaluation of the correlation functions in
Eqs. (\ref{corr2a})-(\ref{corr2c}) the following expressions:
\[
\langle {(\sigma_1\Theta_{r'})}_{jj}{(\sigma_1{\bar\Theta}_{r})}_{jj}\rangle
=-\langle \psi_{r'}{\bar\psi}_r\rangle+o(\eta^{-1})
\]
\[
\langle (Tr_2(\sigma_3\sigma_1\Theta_{r'})Tr_2(\sigma_3\sigma_1{\bar\Theta}_{r})\rangle
=-4\langle \psi_{r'}{\bar\psi}_r\rangle+o(\eta^{-1})
\]
and
\[
\langle Tr_2(-i\sigma_1P_{r'})Tr_2(\sigma_1Q_r)\rangle
=-\eta^2-4\langle\psi_r{\bar\psi}_{r'}\rangle\langle\psi_{r'}{\bar\psi}_r\rangle \ .
\]
This leads to
\beq
C_{rr'}=-\frac{2}{g^2\pi^2}\langle\psi_{r'}{\bar\psi}_r\rangle
=\frac{1}{g^2\pi^2}K^{-1}_{rr'}
\label{corrfin1}
\eeq
and
\beq
A_{rr'}(\omega)=-\frac{4}{g^2}\langle\psi_{r'}{\bar\psi}_r\rangle
=\frac{2}{g^2}K^{-1}_{rr'} \ .
\label{corrfin2}
\eeq
Finally, the correlation function of the local density of states reads 
\beq
D_{rr'}=\frac{1}{g^2\pi^2}(\eta^2+4\langle\psi_r{\bar\psi}_{r'}\rangle\langle\psi_{r'}
{\bar\psi}_r\rangle)
=\frac{\eta^2}{g^2\pi^2}+\frac{1}{g^2\pi^2}K^{-1}_{rr'}K^{-1}_{r'r} \ .
\label{corrfin3}
\eeq
In summary, all the correlations are expressed in terms of the inverse of 
the matrix $K$ of Eq. (\ref{cmatrix}).
Interesting is that only correlations of the Grassmann field $\psi$ appear, whereas the
real fields $\varphi_{jj'}$ do not contribute, at least in the approximation
up to $o(\eta^{-1})$.

\section{Discussion}

All three quantities in Eqs. (\ref{corrfin1}), (\ref{corrfin2}), and (\ref{corrfin3})
are related to the same correlation function, namely to 
$K^{-1}_{rr'}=-2\langle\psi_r{\bar\psi}_{r'}\rangle$ of Eq. (\ref{cmatrix}).
The latter, or its Fourier components $1/K(q)$ with
\beq
K(q)=\frac{8}{\eta^3}[\epsilon'+c(q)],
\hskip0.5cm
\epsilon'=\epsilon-iE 
\ ,
\label{2pp}
\eeq
can be considered as the propagator of the
average two-particle Green's function $\langle G_{rr'}G_{r'r}\rangle$. It describes the motion of
two particles, created at the same time at site $r'$ (cf. Appendix B). Although the 
two particles are independent
(i.e., we ignore their Coulomb interaction here), the averaging over the random vector 
potential creates an {\it effective interaction} between them. This interaction is presented by the
quartic term in $S_1$ of Eq. (\ref{effaction}), whose strength is $g$, the variance of the Gaussian
distributed vector potential. A consequence of the interaction is that the two-particle propagator
describes diffusion, when we consider $\epsilon'=i\omega/2$ and study $q\sim0$:
\[
\frac{1}{K(q)}\sim \frac{\eta^3/4}{i\omega+D q^2}
\]
with the diffusion coefficient
\beq
D=\frac{1}{2}\frac{\partial^2c(q)}{\partial q_l^2}\Big|_{q=0}
=\frac{1}{2\eta}\int_k \sum_j\left(\frac{\partial h_{j}}{\partial k_l}
\frac{\partial h_{j}}{\partial k_l}-\frac{\partial^2 h_{j}}{\partial k_l^2}h_j
\right) \ .
\label{diff}
\eeq
Here it has been assumed that $D$ is isotropic.

Assuming weak disorder, the action $S'$ in Eq. (\ref{spaction}) can be expanded in powers of $\eta$.
(This weak-disorder approach is valid even for $\epsilon=\omega=0$, in contrast to the factorization
approach of Sect. \ref{sectpt})
A diffusion propagator was also found for this case, with a the different diffusion 
coefficient though \cite{ziegler06}:
\[
D_w=\frac{g}{4\pi\eta}  
\]
for Dirac fermions and $E=0$.
Thus, the physics of the average two-particle Green's function
is diffusive, both for weak (i.e. for $\eta\ll 1$) and for strong disorder (i.e. for $\eta\gg1$). 
The diffusion coefficient depends on the
one-particle scattering rate $\eta$ and the Hamiltonian $H_0$, given by its components $h_j$
in Eq. (\ref{ham000}). It is remarkable that
both diffusion coefficients, the one of the strong-disorder expansion in
Eq. (\ref{diff}) as well as $D_w$ of the weak-disorder expansion, are proportional to
the one-particle scattering time $\tau=\eta^{-1}$. 
The latter can be calculated from the self-consistent Born approximation \cite{peres06}.
However, $D$ always decreases with
increasing disorder, whereas $D_w$ is not monotonous with $g$ but has a minimum
due to the extra factor $g$. For realistic values of $g$, where the variance 
of the Gaussian distribution is $0<g<1$, the diffusion coefficient $D_w$ decreases with $g$.
In any case, the increasing behavior of $D_w$ is beyond the validity of the weak-disorder approach.

The integration with respect to the SP manifold of Sect. 5.1 allows us to identify the
relevant transformation parameters for the long-range correlation functions. From the
results in Eqs. (\ref{corrfin1})-(\ref{corrfin3}) it is obvious that only $\psi$, i.e.
the transformation that mixes fermions and bosons, is relevant. On the other hand, 
the transformation inside the bosonic and inside the fermionic sector, provided by
the parameters $\varphi_{jj'}$, is not relevant for the long-range correlations.
This is in agreement with previous calculations for the conductivity, where the
integration with respect to $\varphi_{jj'}$ has not been taken into account \cite{ziegler98}.
If we project the symmetry transformation by choosing $\psi={\bar\psi}=0$ in Eqs. (\ref{u0}) 
and (\ref{u1}), there would also be a Goldstone mode, which becomes massless as we send
the symmetry-breaking term to zero (i.e. $\epsilon\to0$). This case has two interesting
consequences: (i) the average density of states would be divergent at $E=0$ due to
\[
\langle \sigma_1 Q_r\rangle= -i\eta\sigma_0(\langle\varphi_{11,r}^2\rangle -1/2) \ ,
\]
and (ii)
the long-range behavior of $C_{rr'}$, $A_{rr'}$ and $D_{rr'}$ would disappear.

For the calculation of the physical quantities of Sect. 5.1 we need the following expressions.
The diagonal elements of $K^{-1}$ diverge logarithmically with $\epsilon\sim0$ as
\[
K^{-1}_{rr}=\int_q \frac{1}{K(q)}\sim K_0\ln(\epsilon) \ ,
\]
which implies a divergency of the second moment of the local density of states $D_{rr}$.
Summation over the lattice sites gives
\beq
\sum_r K^{-1}_{r0}=\frac{1}{K(q=0)}=\frac{\eta^3}{4\epsilon'} \ .
\label{q1}
\eeq
Finally, for the matrix element of $r_k^2$ and the conductivity we need 
\beq
\sum_r r_k^2K^{-1}_{r0}=-\frac{\partial^2}{\partial q_k^2}\frac{1}{K(q)}\Big|_{q=0}
=\frac{\eta^3}{2\epsilon'^2}D \ .
\label{q2}
\eeq

\subsection{One-particle scattering rate $\eta$}

In the following the one-particle scattering rate $\eta$ will be discussed
for the specific case of Dirac fermions. Then the SP Eq. (\ref{SPE00}) reads
\beq
\eta=\frac{g}{2\pi}(\eta+iE)\ln\left[1+\frac{\lambda^2}{(\eta+iE)^2}\right] \ ,
\label{SPEa}
\eeq
where $\lambda$ is the momentum cutoff of the Dirac fermions. For weak disorder
(i.e. $g\ll 1$), the scattering rate is also weak. To study the Dirac point, we rewrite
Eq. (\ref{SPEa}) as
\beq
\eta=-iE+\frac{\lambda}{\sqrt{\exp(2\pi\eta/g(iE+\eta))-1}} \ .
\label{SPEaa}
\eeq
and take the limit $E=0$
\beq
\eta_0=\frac{\lambda}{\sqrt{e^{2\pi/g}-1}} \ ,
\label{SPEb}
\eeq
This is shown in Fig. 2.
For $|E|\ll\eta_0$ we can solve the SP Eq. (\ref{SPEaa}) by an expansion in $E$:
$\eta=\eta_0+o(E)$. On the other hand, for $|E|\gg|\eta|$ we can iterate Eq. (\ref{SPEa})
with the initial value
\beq
{\bar\eta}=\frac{g}{2\pi}iE\ln\left[1-\lambda^2/E^2\right]
=\frac{gE}{2}\left(sign(E)+\frac{i}{\pi}\ln\left[\lambda^2/E^2-1\right]\right)
\hskip0.5cm
(E^2<\lambda^2) \ .
\label{etai}
\eeq
${\bar\eta}$ is a reasonable approximation of $\eta$ if
\[
g\Big|1+\frac{i}{\pi}\ln\left[\lambda^2/E^2-1\right]\Big|\ll 1 \ .
\]

\subsection{Density of states}

The density of states is proportional to the average one-particle Green's function
\[
\rho(E)=Tr_2\langle G_{rr}(E-i\epsilon)\rangle
\]
and describes two important features of our model, 
the spontaneous symmetry breaking and the one-particle scattering rate $\eta$
in the SP approximation (cf. Eq. (\ref{SPE00}). In terms of the SP integration
of the functional integral,
Eq. (\ref{single}) provides a finite average density of states at the Dirac point
\[
\rho(E=0)=-i\frac{1}{g} Tr_2\langle \sigma_1Q\rangle=
-\frac{1}{g}Tr_2\langle\sigma_1P\rangle=\eta_0/g \ , 
\]
which is practically zero for a larger regime of $g$ (cf. Fig. 2).
Moreover, for $|E|\gg|\eta|$ we get a linear behavior from Eq. (\ref{etai})
\[
\rho(E)\approx\frac{1}{g}Re({\bar\eta})=\frac{|E|}{2} \ ,
\]
which reflects the density of states of pure Dirac fermions. Both results are
in good agreement with a self-consistent calculation of the average density of states
\cite{dora08,hu08}. 

Correlations of the local density of states $D_{rr'}$ have a long-range behavior. The
corresponding Fourier transform
\beq
D(q)=\int_k\frac{1}{K(k-q/2)K(k+q/2)}
\label{ddcorr}
\eeq
is a function of $q^2$ and diverges like $q^{-2}$ at $q=0$. It has
has a kink (or a shoulder) at the edge of the one-particle spectrum $\lambda$ (cf. Fig. 3).

\subsection{Microwave conductivity}

According to Eqs. (\ref{conda1}), (\ref{corrfin2}) and (\ref{q2}), 
the matrix element of $r_k^2$ reads 
\[
\langle\Phi_{\omega/2} |r_k^2|\Phi_{-\omega/2}\rangle
=\sum_r r_k^2 A_{r0}(\omega) 
=-\frac{1}{g^2}\frac{\eta^3}{\omega^2}D\ .
\]
Using the expression of the diffusion coefficient in Eq. (\ref{diff}) and taking $\epsilon\to0$, 
the matrix element becomes
\beq
\langle\Phi_{\omega/2} |r_k^2|\Phi_{-\omega/2}\rangle
=-4\frac{\eta^2}  
{g^2\omega^2}\
\int_k \sum_j\left(\frac{\partial h_{j}}{\partial k_l}
\frac{\partial h_{j}}{\partial k_l}-\frac{\partial^2 h_{j}}{\partial k_l^2}h_j
\right) 
\ .
\label{matrixel1}
\eeq
Lower frequencies $\omega$ (i.e. lower energies) are more important for scattering than 
states of higher energies
due to their larger matrix elements. Moreover, if we interpret the matrix element as a measure of 
localization, the states with $\omega>0$ are localized on a scale $1/\omega$. 
The matrix element, together with Eq. (\ref{conda1}), gives for the microwave conductivity
\beq
\sigma_0(\omega) 
=-\frac{e^2}{4\pi\hbar}\omega^2\langle\Phi_{\omega/2} |r_k^2|\Phi_{-\omega/2}\rangle
\sim\frac{e^2}{\pi\hbar}\frac{\eta_0^2}{g^2}
\int_k \sum_j\left(\frac{\partial h_{j}}{\partial k_l}
\frac{\partial h_{j}}{\partial k_l}-\frac{\partial^2 h_{j}}{\partial k_l^2}h_j
\right) 
\hskip0.5cm
(\omega\ll\eta_0)\ .
\label{condfin}
\eeq
This result indicates a constant microwave conductivity, at least for $\omega\ll\eta_0$,
since the prefactor $\omega^2$ is compensated by the $\omega^{-2}$ behavior of the matrix
element of $r_k^2$. 
It should also be noticed that $\lim_{\omega\to0}\sigma_0(\omega)$ gives the DC conductivity
of Eq. (\ref{cond000}). This follows immediately from the definitions of the two conductivities
in Eqs. (\ref{cond000}), (\ref{conda1}) and from Eqs. (\ref{corrfin1}), (\ref{corrfin2}).

The $\omega^{-2}$ behavior of the $r_k^2$ matrix element in Eq. (\ref{matrixel1}) obviously
does not depend on the special form of $h_j$, as long as the spinor structure of $H_0$ exists.
This implies that also for a parabolic $k_j$ dependence (e.g. in the case of a 
graphene bilayer \cite{katsnelson07c,mccann06,cserti07}), the cancellation of the $\omega^2$ terms in the
conductivity takes place.

\subsection{Participation ratios}

The inverse participation ratio $P^{(2)}$ of Eq. (\ref{ipr1}) vanishes like
\beq
P^{(2)}\sim\epsilon(\ln\epsilon)^2 ,
\label{ipr2}
\eeq
which indicates the existence of delocalized states at the Dirac point. 
The participation ratio
\beq
p^{(2)}\sim (\ln\epsilon)^{-2} \ ,
\label{pr2}
\eeq
on the other hand, vanishes logarithmically. These two results are consistent 
with a critical point at $E=0$, where there is a transition from localized to 
extended states.

\subsection{Perturbation theory for weak disorder and factorization
\label{sectpt}}

The fact that diffusion is controlled by the one-particle scattering rate $\eta$
raises the question about the quality of the one-particle approximation. The latter has
been used frequently by factorizing the two-particle Green's function 
\cite{lee93,koshino06,peres06,castroneto07b}.
This approximation should be valid for weak disorder. It is based on the assumption that
the Green's functions are uncorrelated and the averaged product is approximately
the same as the product of the averaged one-particle Green's functions: 
\[
\langle G^+_{rr',jj}G^+_{r'r,kk}\rangle
\approx \langle G^+_{rr',jj}\rangle\langle G^+_{r'r,kk}\rangle \ .
\]
This allows us to treat the average one-particle Green's functions within the self-consistent
Born approximation:
\[
\langle G^\pm\rangle\approx (H_0 \pm i\eta)^{-1}
\equiv (H_0 \pm i\eta)^{-1} \ ,
\]
where $\eta$ the imaginary part of the self-energy (or inverse scattering time) determined for Dirac 
fermions in Eq. (\ref{SPEa}).
Consequently, the one-particle Green's function decays exponentially on the scale $\eta^{-1}$. 
The correlation function $D_{rr'}$ is constant and proportional to $\eta^2$,
and there is no divergence for $r'=r$.
This means that we have lost in the factorization the correlation term
\[
\frac{1}{g^2\pi^2}K^{-1}_{rr'}K^{-1}_{r'r}
\]
of Eq. (\ref{corrfin3}). These are the substantial differences between the strong-disorder
expansion and weak-disorder perturbation theory at the Dirac point.
If we go away from the Dirac point, we can study the matrix element $r_k^2$,
approximated by the factorization as
\[
\langle\Phi_{\omega/2} |r_k^2|\Phi_{-\omega/2}\rangle
=-\sum_r r_k^2 Tr_2[\sigma_3G_{r0}(\omega/2+i\eta)\sigma_3G_{0r}(\omega/2+i\eta)]
\sim\cases{
-\eta_0^{-2} & for $\eta_0\gg\omega$ \cr
-4\omega^{-2} & for $\eta_0\ll\omega$ \cr
} \ ,
\]
where we have assumed that $\omega\ll\lambda$.
In contrast to the result of Eq. (\ref{matrixel1}), the matrix element does not diverge now
if we approach the Dirac point $\omega=0$, since there is the finite limit
$\eta_0^{-2}$. This reflects the finite decay length $\eta_0^{-1}$ of the average one-particle
Green's function. Away from the Dirac point the factorization works better.
This can be seen if we insert the matrix element of $r_k^2$ into the conductivity
of Eq. (\ref{conda1})
\beq
\sigma_0(\omega)
=-\frac{e^2}{4\pi\hbar}\omega^2\langle\Phi_{\omega/2} |r_k^2|\Phi_{-\omega/2}\rangle
\sim \frac{e^2}{\pi\hbar} =2\frac{e^2}{h}
\hskip0.5cm
(\eta_0\ll \omega)
\eeq
in the perturbative regime. Thus again, the conductivity does not depend on the frequency.
Actually, the validity of the perturbative regime for the matrix element of $r_k^2$ and the
conductivity is big in terms of $g$ due to $\eta_0\sim e^{-\pi/g}$.

\section{Conclusions}

Spinor states, described by a two-dimensional Dirac-like lattice Hamiltonian, were studied in an uncorrelated 
random vector potential. Our calculation, based on a strong-disorder expansion, has revealed that the quantum states 
develop long-range correlated fluctuations. In other words, the quantum system transforms the uncorrelated 
fluctuations of the random vector potential into long-range correlated fluctuations, for instance, of the 
density of states. The origin of this behavior is spontaneous symmetry breaking, which develops a massless 
(long-range) mode. The spontaneous symmetry breaking is measured by the one-particle scattering rate $\eta$ or 
the density of states $\eta/g$. An important scale for this effect is
\[
\eta_0\sim e^{-\pi/g} \ ,
\]
which depends on the variance of the Gaussian fluctuations of the random vector potential g. 
(Although our approach is not valid for very small $g$ (cf. Sect. 6.5), this scale is short for all 
reasonable values $g\approx 1$.) $\eta_0$ separates regimes that are controlled by disorder ($\eta\ll\omega$) 
from that which is controlled by energy (or frequency) ($\eta\ll\omega$).
For instance, a central quantity is the one-particle scattering rate has a crossover with respect to
$\omega$ as
\[
\eta\sim\cases{
e^{-\pi/g} & for $\eta_0\gg\omega$ \cr
\omega g & for $\eta_0\ll\omega$ \cr
} \ .
\]
Perturbation theory with respect to disorder can be applied to the regime with $\eta_0\ll\omega$, 
at least for quantities like the conductivity. The correlation length is not affected qualitatively by 
this cross over, as one can see from the matrix elements of $r_k^2$
\[
\langle\Phi_{\omega/2} |r_k^2|\Phi_{-\omega/2}\rangle
=-4\omega^{-2}
\cases{
(\eta_0^2/g^2)  \int_k \sum_j\left(\frac{\partial h_{j}}{\partial k_l}
\frac{\partial h_{j}}{\partial k_l}-\frac{\partial^2 h_{j}}{\partial k_l^2}h_j
\right) & for $\eta_0\gg\omega$ \cr
1 & for $\eta_0\ll\omega$ \cr
} \ ,
\]
which provides an effective correlation length $\omega^{-1}$ for $\langle\Phi_{E,rj}\Phi^*{-E,rj}\rangle$.
As we go away from the Dirac point, the correlation length decreases. 
On the other hand, the correlation function of the local density of states $D_{rr'}$
clearly distinguishes both regimes, since the corresponding Fourier transform is
\[
D(q)\sim\cases{
D_0q^{-2} & for $q\sim0$ and $\eta_0\gg\omega$ \cr
D_1\delta(q) & for $\eta_0\ll\omega$ \cr
} \ .
\]
An interesting consequence of the "universal" $\omega^{-2}$ behavior of the matrix element of 
$r_k^2$ is a constant microwave conductivity in both regimes:
\[
\sigma_0(\omega)\sim 2\frac{e^2}{h}\cases{
(\eta_0^2/g^2)
\int_k \sum_j\left(\frac{\partial h_{j}}{\partial k_l}
\frac{\partial h_{j}}{\partial k_l}-\frac{\partial^2 h_{j}}{\partial k_l^2}h_j
\right) & for $\eta_0\gg\omega$ \cr 
1 & for $\eta_0\ll\omega$ \cr
} \ .
\]
This behavior must be seen in contrast to the conventional Drude behavior, 
where the real part of conductivity decays like $\omega^{-2}$. It is a consequence
of the correlated scattering between the upper and the lower part of the Dirac cones
(i.e., it is a manifestation of the zitterbewegung). This effect should be experimentally observable,
since it survives also in the presence of disorder. 

Our results can be summarized by the statement that properties of graphene at the Dirac point
are ruled by long-range correlations. At the distance $\omega$ from the Dirac point they decay 
on the scale $\omega^{-1}$. This has physical consequences, like a constant microwave
conductivity, which also has been observed experimentally \cite{nair08}. Disorder of
variance $g$ creates a characteristic scale $\eta_0=e^{-\pi/g}$, which separates the
behavior in a vicinity $\omega\ll \eta_0$ of the Dirac point from another one away
from the Dirac point with $\omega\gg \eta_0$. Standard perturbation theories and
simple approximations can be applied to the latter because then the behavior is ruled
by one-particle properties.  

\vskip0.5cm

\no
Acknowledgement:

\no
I am grateful to B. D\'ora for useful discussions.

\section*{Appendix A: Conductivity}

The conductivity per site on a lattice with $N$ sites can be evaluated within the Kubo formalism and
gives (cf. Eq. (3) in Ref. \cite{ziegler06})
\[
Re(\sigma_{kk})=\frac{e^2}{N\hbar}\pi\int Tr\left[[H,r_k]\delta(H-E+\omega)[H,r_k]
\delta(H-E)\right]\frac{f_\beta(E+\omega)-f_\beta(E)}{\omega}dE
\]
\[
=\frac{e^2}{N\hbar}\pi\omega^2\int Tr\left[r_k\delta(H-E+\omega/2)r_k
\delta(H-E-\omega/2)\right]
\frac{f_\beta(E+\omega/2)-f_\beta(E-\omega/2)}{\omega}dE
\]
with the Fermi function at inverse temperature $\beta$: $f_\beta(E)=1/(1+\exp(\beta E))$.
Now we consider
\[
Tr\left[r_k\delta(H-E+\omega/2)r_k\delta(H-E-\omega/2)\right]
=\sum_{r,r'}r_kr_k' Tr_2\left[\delta_{rr'}(H-E+\omega/2)
\delta_{r'r}(H-E-\omega/2)\right]
\]
We can write
\[
2r_kr_k' =-(r_k-r_k')^2+r_k^2+{r_k'}^2
\]
Moreover, $\delta(H-E+\omega/2)\delta(H-E-\omega/2)=0$ for $\omega\ne0$.
Thus we obtain 
\[
\sum_{r,r'}r_kr_k' Tr_2\left[\delta_{rr'}(H-E+\omega/2)\delta_{r'r}(H-E-\omega/2)\right]
\]
\[
=-\frac{1}{2}\sum_{r,r'}(r_k-r_k')^2 Tr_2\left[\delta_{rr'}(H-E+\omega/2)
\delta_{r'r}(H-E-\omega/2)\right]
\]
This gives
\[
-\frac{e^2}{2N\hbar}\pi\omega^2\int 
\sum_{r,r'}(r_k-r_k')^2 Tr_2\left[\delta_{rr'}(H-E+\omega/2)
\delta_{r'r}(H-E-\omega/2)\right]
\frac{f_\beta(E+\omega/2)-f_\beta(E-\omega/2)}{\omega}dE \ .
\]
Since for low temperatures 
\[
f_\beta(E)\sim\Theta(-E) \ ,
\]
we get
\[
\sim \frac{e^2}{2\hbar}\pi\omega
\sum_{r}(r_k-r_k')^2 \int_{-\omega/2}^{\omega/2}Tr_2\left[\delta_{rr'}(H-E+\omega/2)
\delta_{r'r}(H-E-\omega/2)\right]dE
\]
The Dirac delta functions can be expressed by one-particle Green's functions:
\[
\delta(H+\omega/2)=\frac{1}{2\pi i}[G(\omega/2-i\epsilon)-G(\omega/2+i\epsilon)]
\]
such that
\[
Re(\sigma_{kk})\approx \sigma^I(\omega)+\sigma^{II}(\omega)
\]
with
\[
\sigma^I(\omega)
=\frac{e^2}{4h}\omega\int_{-\omega/2}^{\omega/2}
\sum_{r}(r_k-r_k')^2 Tr_2[
G_{rr'}(\omega/2-E-i\epsilon)G_{r'r}(-\omega/2-E-i\epsilon)
\]
\[
+G_{rr'}(\omega/2-E+i\epsilon)G_{r'r}(-\omega/2-E+i\epsilon)]dE
\]
and
\[
\sigma^{II}(\omega)
=-\frac{e^2}{4h}\omega\int_{-\omega/2}^{\omega/2}
\sum_{r}(r_k-r_k')^2 Tr_2[G_{rr'}(\omega/2-E-i\epsilon)G_{r'r}(-\omega/2-E+i\epsilon)
\]
\[
+G_{rr'}(\omega/2-E+i\epsilon)G_{r'r}(-\omega/2-E-i\epsilon)]dE 
\]
and because of $G(z)=-\sigma_3G(-z)\sigma_3$ is
\[
\sigma^{II}(\omega)
=-\frac{e^2}{4h}\omega\int_{-\omega/2}^{\omega/2}
\sum_{r}(r_k-r_k')^2 Tr_2[G_{rr'}(\omega/2-E-i\epsilon)G_{r'r}(-\omega/2-E+i\epsilon)
\]
\[
+G_{rr'}(-\omega/2+E-i\epsilon)G_{r'r}(\omega/2+E+i\epsilon)]dE \ .
\]

\section*{Appendix B: matrix elements of the energy states}

Starting from a (localized) state $\Psi(0)$, we can allow the state to evolve in time:
\[
\Psi(t)=e^{-iHt}\Psi(0)
\]
The question is how the states $\Psi_{\pm E}$ can be reached by this evolution, and
how this is influenced by scattering due to disorder.
The contribution of the state with energy $\pm E$ to the time evolution is obtained 
by the Fourier transformation $t\to \pm E$ of $\Psi(t)$ for positive time $t\ge 0$ (since
the wave function did not exist for $t<0$):
\[
\Phi_{\pm E}\equiv\int_0^\infty e^{(\pm iE-\epsilon)t}\Psi(t)dt
=\int_0^\infty e^{(\pm iE-\epsilon)t}e^{-iHt}dt\Psi(0)
\]
\[
=-i(H \mp E-i\epsilon)^{-1}\Psi(0)=-iG(\mp E-i\epsilon)\Psi(0) \ .
\]
Since $H$ is Hermitean (i.e. $H^\dagger=H$), the complex conjugate of the wave function is
\[
\Phi_{\pm E}^*=iG(\mp E-i\epsilon)^*\Psi^*(0) =iG(\mp E+i\epsilon)^T\Psi^*(0)\ ,
\]
where $G^T$ is the transposed of $G$. The matrix element of $r_k^2$ between $\Phi_{\pm E}$ is
\beq
\langle\Phi_{E}|r_k^2|\Phi_{-E}\rangle
=\sum_{r}r_k^2\Phi_{E,rj}\Phi_{-E,rj}^*
=\sum_{r,j'} r_k^2 G_{r0,jj'}(-E-i\epsilon)G_{0r,j'j}(E+i\epsilon)|\Psi_{j'}(0)|^2
\label{overlap0}
\eeq
if we assume that $\Psi(0)$ is localized at the origin of the lattice $r=0$. 
In the presence of disorder, this expression should be averaged with respect to the latter:
\beq
\langle\Phi_{E}|r_k^2|\Phi_{-E}\rangle
=\sum_{r,j}r_k^2\langle\Phi_{E,rj}\Phi_{-E,rj}^*\rangle 
=\sum_{r,j,j'} r_k^2\langle G_{r0,jj'}(-E-i\epsilon)G_{0r,j'j}(E+i\epsilon)\rangle |\Psi_{j'}(0)|^2
 \ .
\label{corr00a}
\eeq
It describes a correlation between the states in the upper and in the lower band,
if they evolve from the same initial state $\Psi(0)$.
These results can be summarized to the relation
\[
\sum_{r} r_k^2\langle Tr_2\left[
G_{r0}(-E-i\epsilon)G_{0r}(-E'+i\epsilon)\right]\rangle 
=\langle\Phi_{E}|r_k^2|\Phi_{E'}\rangle
\]
if $|\Psi_{j'}(0)|^2=1$ for $j'=1,2$.

\section*{Appendix C: Inverse participation ratios}

The inverse participation ratio is related to the fourth moment of the normalized eigenfunction
$\Psi_{k}$ with eigenvalue $E_k$ as
\beq
P^{(2)}=\langle\sum_k|\Psi_{k,r}|^4\delta(E-E_k)\rangle \ .
\label{ipra}
\eeq
This expression vanishes in the delocalized regime. The latter is plausible when we estimate the
delocalized wavefunction by $|\Psi_{k,r}|^2\sim 1/N$ on a lattice with $N$ sites, taking into
account normalization
\[
\sum_r|\Psi_{k,r}|^2=1 \ .
\]
This implies
\[
\sum_r\sum_k|\Psi_{k,r}|^2\delta(E-E_k)\sim\sum_k\delta(E-E_k)\sim N\rho \ ,
\]
where $\rho$ is the spatially averaged density of states. Moreover, for $P^{(2)}$
we get from Eq. (\ref{ipra}) a vanishing expression for $N\sim\infty$:
\[
\sum_k|\Psi_{k,r}|^4\delta(E-E_k)\sim\frac{1}{N^2}\sum_k\delta(E-E_k)\sim \frac{\rho}{N} \ .
\]
$P^{(2)}$ can also be related to $D_{rr}$ by using the relation \cite{wegner80}
\[
\lim_{\epsilon\to0}\epsilon\sum_k|\Psi_{k,r}|^2\delta_\epsilon(E-E_k)\sum_{k'}|\Psi_{k',r}|^2\delta_\epsilon(E-E_{k'})
=\sum_k|\Psi_{k,r}|^4\delta(E-E_k)
\] 
which implies
\[
P^{(2)}=\lim_{\epsilon\to0}\epsilon D_{rr} \ .
\]

\section*{Appendix D: Random-walk expansion}

The presence of the Grassmann field in the action $S'$ of Eq. (\ref{spaction})
enables us to write the functional integral $\int\exp(-S'){\cal D}[\psi,\varphi]$ in terms of a
dense system of self-avoiding random walks (or polymers), whose density is controlled
by the symmetry-breaking term. Starting from $S'$ we either use $UU'$ of Eq. (\ref{fullsymm})
or its approximation given in Eq. (\ref{approxfield}). In both cases the structure of the action 
with symmetry-breaking parameter $\epsilon$ is
\[
S'=S_0+\epsilon\sum_r\psi_r{\bar\psi}_r +\sum_{r,r'}a_{rr'}\psi_r{\bar\psi}_{r'}
+\sum_{r,r'}b_{rr'}\psi_r{\bar\psi}_r\psi_{r'}{\bar\psi}_{r'} \ .
\]
$S_0$ is a term without Grassmann field.
The quartic term (which does not appear in the approximated field of Eq. (\ref{approxfield}))
can be expressed by a quadratic term that couples to a random field. Thus the general structure
in terms of the Grassmann field reads
\[
S'=S_0+\epsilon\sum_r\psi_r{\bar\psi}_r +\sum_{r,r'}B_{rr'}\psi_r{\bar\psi}_{r'} \ ,
\]
where the coefficients $B_{rr'}$ are random and connect only nearest-neighbor sites (this is a consequence
of the symmetry). The integration
with respect to $\psi$ can be performed first and gives just a determinant
\[
\int\exp(-S'){\cal D}[\psi] = e^{-S_0}det(\epsilon + B) \ .
\]
Then the determinant can be expanded with respect to $B$, which creates random walks, whereas
$\epsilon$ creates an environment of identical points. In other words, a lattice site $r$ is
either occupied by an $\epsilon$ or it is visited by a random walk with jump rate
$B_{rr'}$ between $r$ and a nearest-neighbor site $r'$. The random walks do not intersect
themselves or each other and must be closed. This is a consequence of the integral over the Grassmann field
or equivalently it is determined by the structure of the determinant. The contributions of the
random walks depend on the weight of its elements $B_{rr'}$, relative to the value of 
$\epsilon$. For $\epsilon\sim 0$ (i.e. near the Dirac point) the lattice is completely
covered by closed self-avoiding random walks, which can also be considered as the regime of
dense polymers. Next, we need to integrate with respect to $\varphi$. In general, this will
affect the weight of the random walks. In the special case of $S'$ in Eq. (\ref{faction}) we obtain
\beq
\int\exp(-S'){\cal D}[\psi,\varphi]=\int e^{-S_0}det(K(\epsilon-iE)+zK(0)z){\cal D}[\varphi] \ ,
\label{determinantk}
\eeq
where $z$ is a diagonal matrix with matrix elements $z_{rr}=(\varphi_{11,r}+ \varphi_{22,r})/\eta$
and
\[
S_0=\sum_{r,r'}K_{rr'}(\epsilon-iE)
(-\varphi_{11,r}\varphi_{11,r'}+\varphi_{22,r}\varphi_{22,r'}) \ .
\]
The expansion of the determinant produces terms with products 
$\prod_r z_{rr}^{l_r}=\prod_r[(\varphi_{11,r}+ \varphi_{22,r})/\eta)]^{l_r}$ ($l_r=0,1,2$), where $l_r>0$ 
is created by the
second term in its argument $z_{rr}K_{rr'}(0)z_{r'r'}$. Integration with the weight factor $e^{-S_0}$, where
at least one site $r$ appears with $l_r>0$, gives a vanishing Gaussian integral
\[
\int e^{-S_0} \prod_r(\varphi_{11,r}+ \varphi_{22,r})^{l_r}=0 \ ,
\]
since the two quadratic terms in $S_0$ appear with opposite signs (i.e. $\varphi_{11}$ must be integrated along
the imaginary axis). Therefore, the second term in the determinant of Eq. (\ref{determinantk}) does not
contribute to the functional integral.


\begin{figure}
\centering
\includegraphics[width=0.2\textwidth]{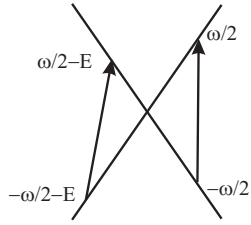}
\caption{Scattering on the Dirac cone: schematic scattering process between energy states
$\Phi_{E-\omega/2}$ and $\Phi_{E+\omega/2}$ as well as between $\Phi_{-\omega/2}$ and $\Phi_{\omega/2}$.
This type of scattering is relevant for the microwave conductivity $\sigma^{II}(\omega)$ in Eq. (\ref{cond0b}).}
\label{plotconda1}
\end{figure}

\begin{figure}
\centering
\includegraphics[width=0.5\textwidth]{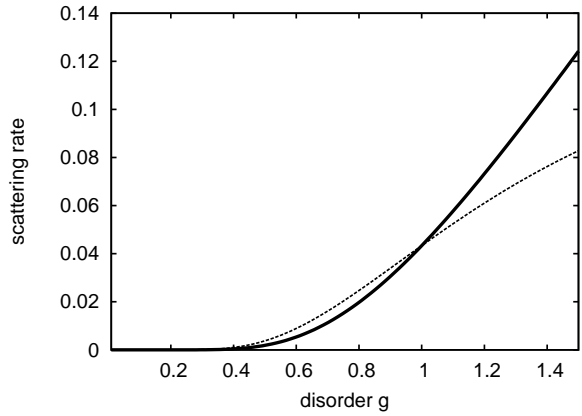}
\caption{Scattering rate (full curve) and density of states at the Dirac point (dashed curve) 
for Dirac fermions. Both quantities practically vanish over a wide range of disorder $g$.}
\label{plotconda2}
\end{figure}

\begin{figure}
\centering
\includegraphics[width=0.5\textwidth]{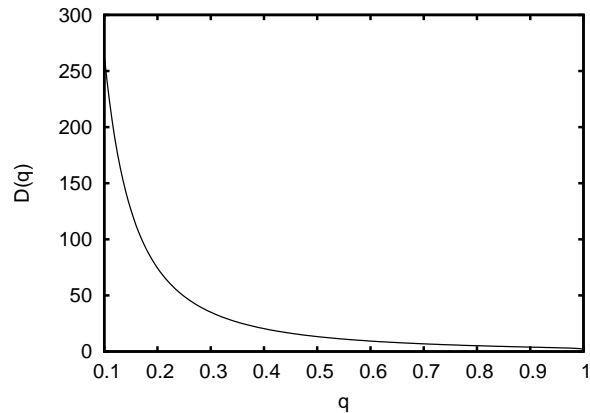}
\caption{Fourier transform of the local density of states correlation function, which diverges like $q^{-2}$.}
\label{plotconda}
\end{figure}

\end{document}